\documentclass{aa}  
\usepackage{graphicx}
\usepackage{txfonts}
\usepackage{hyperref}

\usepackage{xcolor}
\usepackage[normalem]{ulem}
\usepackage{orcidlink}

\begin{document} 

\title{The impact of the white dwarf initial-final mass relation on star clusters' ages inferred from their cooling sequence} 

\author{
M.\,Salaris\inst{1,2}\orcidlink{0000-0002-2744-1928}\fnmsep\thanks{\email{maurizio.salaris@inaf.it}}, S.\,Cassisi\inst{3,4}, L.\,R.\,Bedin\inst{5}\orcidlink{0000-0003-4080-6466}
%\and 
%L.\,R.\,Bedin\inst{4}\orcidlink{0000-0003-4080-6466}
}

\institute{
INAF -- Osservatorio di Astrofisica e Scienza dello Spazio di Bologna, via Gobetti 93/3, I-40129 Bologna, Italy
\and
Astrophysics Research Institute, Liverpool John Moores University, 146 Brownlow Hill, Liverpool L3 5RF, UK
\and
INAF -- Osservatorio Astronomico di Abruzzo, via M. Maggini snc, I-64100 Teramo, Italy
\and
INFN -- Sezione di Pisa, Largo Pontecorvo 3, 56127 Pisa, Italy
\and
INAF -- Osservatorio Astronomico di Padova, Vicolo dell'Osservatorio 5, Padova I-35122, Italy
}

\date{XXX,YYY,ZZZ}
 
\abstract
{
We have investigated and quantified how the choice of the white dwarf (WD) initial-final mass relation (IFMRs) affects the age determination of star clusters from the luminosity function of their cooling sequences.  We have performed a purely theoretical differential analysis using recent semi-empirical IFMRs across three age regimes: $\approx$10\,Gyr (old ages), $\approx$1\,Gyr (intermediate ages), and $\approx$100\,Myr (young ages), respectively. For each regime, we have considered the age of a representative cluster whose entire WD sequence has already been observed and analysed, and calculated theoretical luminosity functions (varying the IFMR) that include realistic observational errors and binning.
We have found that for old ages (as those of globular clusters), the choice of the semi-empirical IFMR introduces age offsets of at most $\sim$0.6$\pm$0.2\,Gyr, while for intermediate ages the impact is reduced to $\sim$0.2$\pm$0.1~Gyr, and it becomes negligible at young ages. Additionally, we have employed the metallicity-dependent IFMR derived by a recent grid of theoretical stellar evolution models to study whether the predicted metallicity dependence impacts the ages of old clusters with subsolar metal content. We found that neglecting this predicted metallicity dependence in old metal-poor globular clusters can lead to an age underestimate of up to 0.8$\pm$0.2\,Gyr. All these age offsets should be interpreted as a systematic uncertainty associated with the choice of the IFMR.
}


\titlerunning{The white dwarf IFMR and star clusters' ages}
\authorrunning{M.\,Salaris et al.}
\maketitle

\section{Introduction}\label{Section1}

The initial–final mass relation (IFMR) of low- and intermediate-mass stars is a fundamental benchmark with profound implications across various astrophysical contexts. By mapping the initial main sequence (MS) mass of a star to its remnant white dwarf (WD) mass, the IFMR provides a direct measure of the integrated mass loss experienced throughout the star's lifetime, which, in turn, is a main ingredient to calculate, for example, the chemical evolution history and the mass-to-light ratio of the host stellar population. From a theoretical stellar evolution point of view, the IFMR sets a lower limit for the initial mass of single stars exploding as supernovae, and it is also necessary to calculate WD isochrones and luminosity functions to   determine the age of
stellar populations. 

It is difficult to make solid theoretical predictions for the IFMR because of our poor knowledge of the efficiency
of mass-loss processes, and the uncertainties in the 
predicted size of CO cores
during the Asymptotic Giant Branch (AGB) evolution, resulting
mainly from uncertainties in the treatment of the thermal pulse
phase \citep[see, e.g.,][]{ir,lk}. Therefore, starting with \citet{weidemann}, several  (semi-)empirical approaches have been devised to determine the IFMR without using AGB models. The most employed and direct (with minimum assumptions) technique is based on observations of bright WDs in star clusters \citep[see, e.g., the detailed analysis by][]{salarismimf}. In a nutshell, a comparison of the cooling ages of a cluster's WDs with known mass, with the MS turn-off cluster's age, provides the necessary information to infer the IFMR of the observed WD sample.
Other techniques use WD-MS, WD-subgiant, WD-WD pairs in wide binaries \citep[e.g.][]{catalanmimf,bcmimf,andrmimf}, where both components are assumed coeval and expected to have evolved without interactions, while recently local field WDs have also been employed to determine the IFMR \citep[][]{elbadrymimf,cunnimf}; in this latter case, among other parameters, the star formation history of the host population needs to be known. These techniques have led in the last quarter of a century to several determinations of the IFMR
\citep[e.g.][]{weidmimf,ferrariomimf,catalanmimf,Kaliraimimf,salarismimf,williamsmimf,dobbiemimf,zhaomimf,gesickimimf,andrmimf,cummimf,elbadrymimf,mcmimf,bcmimf,cunnimf,millermimf}, often inconsistent with each other.

As mentioned before, the IFMR is an ingredient that enters the calculation of WD isochrones and luminosity functions (LFs), 
the tools we employ to determine the age of globular \citep[see, e.g.][]{hansenm4,hansen6397,M4wd,hansen47tuc,gb,N6752hst,
omegacenwdjwst,n6397jwst,M4wdjwst} and open \citep[see, e.g.,][]{n2477wd,richerm67,N2420wd,m37kalirai,N6791wd,N2158,M67lbt,N6819,M37wdgaia} clusters from their observed WD cooling sequences, independently of the MS turn off. 

In fact, different points (denoted by a pair of brightness and colour values) along an isochrone of age $t$ are populated by WD models of different masses but the same total age $t$, equal to the cooling age at that point, plus the corresponding lifetime of the progenitor, whose mass is determined by the IFMR. 
The brightness of the faint end of an isochrone depends on $t$, hence it is the age-indicator when studying cooling sequences of star clusters; older WD isochrones (larger $t$) reach fainter magnitudes because of the longer cooling times of any fixed mass, given that the progenitor lifetimes are constants. 

The actual value of the magnitude reached by a given WD mass at an isochrone age $t$ does depend on 
the progenitor lifetime, hence on the chosen IFMR. Also, given the distribution of the WD and the corresponding progenitor masses along the isochrone, by assuming a mass function for the progenitors, it is possible to predict theoretical WD counts as a function of their brightness (the LF), which are therefore affected by the IFMR. 
For these reasons, variations of the adopted IFMR can, in principle,  affect the isochrone and LF shape, the magnitude range spanned at fixed age, and the mass distribution as a function of luminosity \citep[these are the properties employed to constrain the IFMR from the study of the local field WD population, see][]{elbadrymimf,cunnimf}. However, only a few studies have presented preliminary tests of how star clusters' ages derived from their WD cooling sequences are affected when different IFMRs are chosen 
\citep[see, e.g.,][]{N6752hst,M37wdgaia}, generally finding a negligible variation.

This paper more comprehensively investigates the influence of the adopted IFMR on the theoretical WD LFs used to determine the ages of star clusters. 
We have employed several IFMRs among the semi-empirical 
determinations listed above, and also a set of theoretical IFMR predictions, to assess how different choices affect both  
theoretical isochrones and LFs, and their impact on age determinations. Isochrones and LFs have been used differentially for population ages on the order 
of 100~Myr, 1~Gyr and 10~Gyr, respectively; for a fixed age of the LF calculated with a chosen reference IFMR, we have investigated by how much the ages of the LFs calculated with the other IFMRs have to be varied to match the magnitude of the age-sensitive features of the reference LF. 
For each of the three age regimes, the LFs include photometric errors and binning of representative observed clusters from previous studies. These age variations (if any) should be 
seen as systematic errors in the ages of star clusters 
derived from their cooling sequences.

This kind of differential analysis minimises the effects of potential inconsistencies arising from the use of semi-empirical or theoretical IFMRs in determining WD stellar population ages. Semi-empirical IFMRs are derived using stellar lifetimes for the WD progenitors and WD cooling ages from models that may differ (in terms of evolutionary times) from the progenitor and WD models used to calculate the isochrones and LF for the actual stellar population study. This could cause inconsistencies in the absolute age scale of the LFs\footnote{The same is true when 
using theoretical IFMRs --based on stellar evolution calculations from the main sequence until the beginning of the final WD stage-- but progenitor lifetimes for the calculation of isochrones and LFs which come from different sets of stellar models.}.

The paper is structured as follows: Section~2 describes the adopted 
set of IFMRs and the theoretical WD models and isochrones, while 
Section~3 analyses the effect of the selected IFMRs on the LFs and the implications for WD age determinations in the above-mentioned age regimes. 
A summary and discussion close the paper. 

\section{The adopted IFMRs and WD models}\label{S2}

\begin{figure}
 \includegraphics[width=\columnwidth]{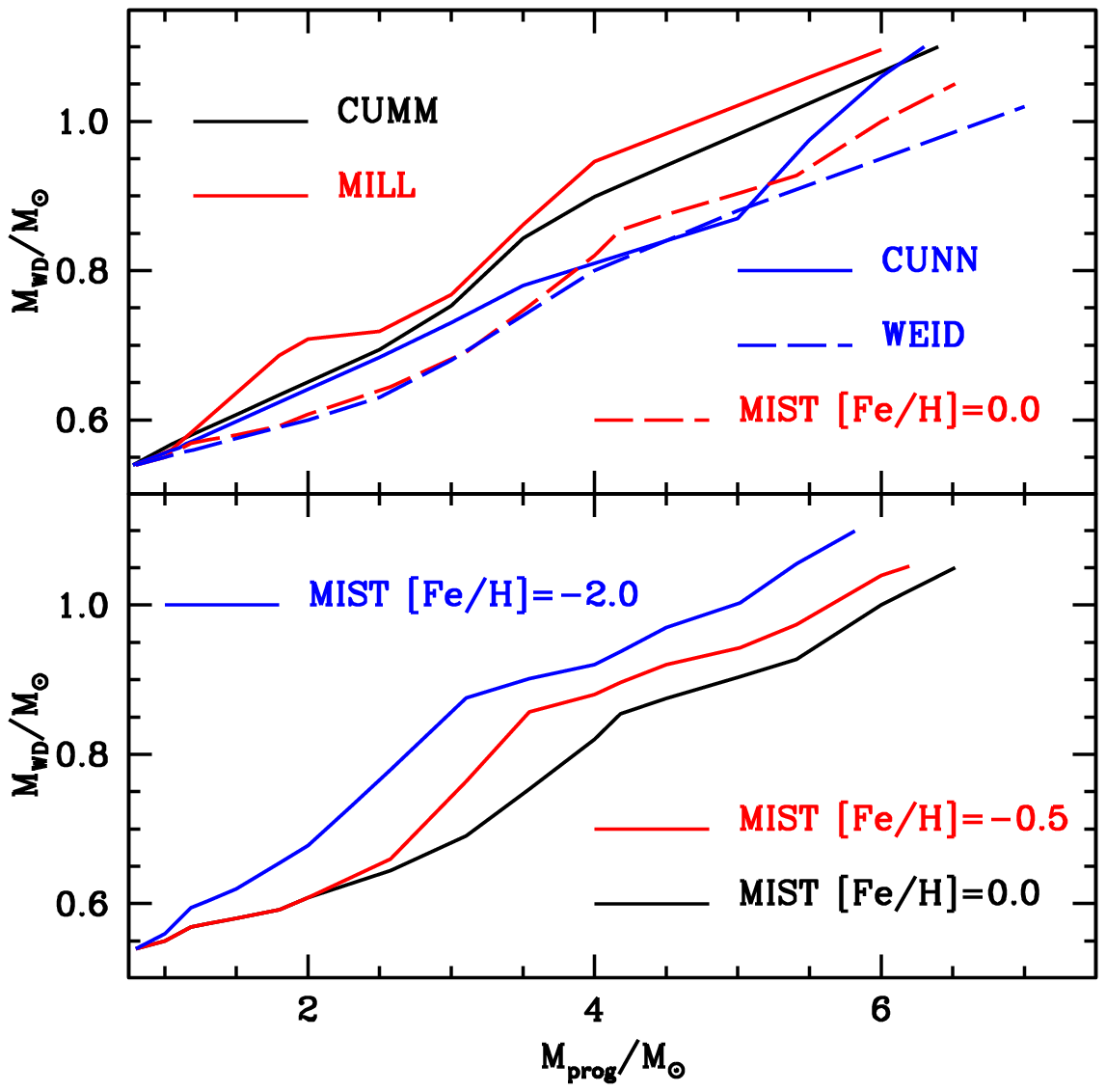}
 \caption{Comparison of the set of semi-empirical and theoretical IFMRs adopted in this study (see text for details); the vertical axis 
 displays the WD final mass ($M_{\rm WD}$), whilst the horizontal axis shows the initial progenitor mass ($M_{\rm prog}$), both in solar units.
 {\sl Upper panel:} 
MILL (solid red), CUMM (solid black), CUNN (solid blue), WEID (dashed blue) semi-empirical IFMRs, and the theoretical MIST results for [Fe/H]=0 (dashed red).
{\sl Lower panel:} Theoretical MIST IFMRs for [Fe/H]=0 (solid black -- this same IFMR is also plotted as a dashed red line in the upper panel), 
[Fe/H]=$-$0.5 (solid red) and [Fe/H]=$-$2.0 (solid blue).}
\label{ifmr} 
\end{figure}

We have employed the BaSTI-IAC progenitor and the CO-core\footnote{The set computed with the electron conduction opacities by \citet{pot}}, hydrogen envelope WD models \citep{bastiss, bastiaen, bastiwd}\footnote{Models are available at \url{http://basti-iac.oa-abruzzo.inaf.it/}}.  
With these inputs, we have computed sets of WD isochrones by employing the IFMRs displayed in Fig.~\ref{ifmr}. To calculate the corresponding LFs, we have assumed as reference a power law for the number distribution at birth of the WD progenitor masses, with a Salpeter exponent $k=-$2.35. This seems an appropriate 
reference choice for the mass range of the progenitors of the WDs we see today in 
open and globular clusters \citep[see, e.g.,][]{bastian}.

The IFMRs shown in Fig.~\ref{ifmr} (also listed in  Table~\ref{tab}) are the semi-empirical results by Weidemann (\citealt{weidmimf} -- hereafter WEID), Cummings (\citealt{cummimf} -- CUMM), Cunningham (\citealt{cunnimf} -- CUNN), and Miller (\citealt{millermimf} -- MILL). If the adopted IFMR reaches WD masses above 1.1$M_{\odot}$ --the largest mass of the BaSTI-IAC CO-core WD models-- as in the case of 
the MILL and CUNN results, we have neglected the part of the IFMR above this limit. In Appendix~A, we briefly discuss the impact on our analysis of including more massive ONe-core WDs with masses above 1.1$M_{\odot}$.

\begin{table}
\centering
\caption{\label{tab} IFMRs adopted in this study.}
{
\begin{tabular}{lcll}
\hline
Acronym & References & Method & Notes \\
\hline

CUMM   & 1  &  clusters' WDs  &  PARSEC\\
    &                     &  (semi-empirical) & progenitor\\
     &                     &  & lifetimes\\
\hline
MILL   & 2 &  clusters' WDs  &  fit 2\\
    &                     &  (semi-empirical) &\\
\hline
CUNN   & 3  &  field WDs  &  \\
    &                     &  (semi-empirical) &\\
\hline
WEID & 4 & clusters' WDs &   for heuristic\\
 & & (semi-empirical) & purposes only\\
 \hline
MIST   & 5 &  stellar models  &  solar scaled \\
& & & progenitors\\
\hline
\end{tabular}
\tablebib{
(1)~\citet{cummimf}; (2) \citet{millermimf}; (3) \citet{cunnimf}; (4) \citet{weidmimf};
(5) \citet{bauermesa}.
}
}
\end{table}

In the case of the MILL IFMR, we have used their fit~2 \citep[see][]{millermimf}, which is in better agreement with the recent analysis by \citet{abia} regarding the IFMR in the range of progenitor masses ($ M_{\rm prog}$) between $\sim$1.5 and $\sim$ 2$M_{\odot}$ \citep[see the discussion in][]{abia}.
As for the CUMM IFMR, we have employed the results obtained by \citet{cummimf} using the PARSEC \citep{parsec} models for the progenitor lifetimes. \citet{cummimf} also provides an alternative IFMR derived using \citet{choi} models for the progenitor lifetimes; this latter IFMR, however, differs from our adopted CUMM IFMR by much less than any other IFMR considered in this analysis.

The MILL result is the most recent IFMR based on WDs in star clusters, whilst the CUNN IFMR is the latest determination based on local field WDs. The CUMM IFMR is a widely used relation also based on WDs in star clusters, which we have employed in our more recent works on WD cluster ages \citep[e.g.][and see later in this section]{M37wdgaia,n6397jwst,omegacenwdjwst,M4wdjwst,47Tucwdjwst}.
Finally, the WEID IFMR is the \lq{classic\rq} result based on earlier studies of clusters' WDs, superseded by the more updated CUMM and MILL results. 
We have selected it for heuristic purposes only, because it is the most discrepant IFMR amongst the set in Fig.~\ref{ifmr}, and has allowed us in Sect.~\ref{S3} to maximise and better explain the effect of IFMR variations on isochrones and LFs. We will refer only to the CUMM, MILL and CUNN IFMRs when we summarise our results in Sect.~\ref{S4}

For a given value of $ M_{\rm prog}$, these IFMRs roughly cover the entire range of final WD masses ($M_{\rm WD}$) spanned by the full set of semi-empirical determinations listed in Sect.~\ref{Section1}.
Across the entire range of initial masses, the WEID IFMR generally gives the lowest final $M_{\rm WD}$ values, whilst the MILL IFMR provides the highest $M_{\rm WD}$. To give a quantitative idea 
of the differences at fixed $M_{\rm prog}$, for progenitor masses between $\sim$4 and $\sim 6 M_{\odot}$ the full range of $M_{\rm WD}$ spanned by the semi-empirical IFMRs in Fig.~\ref{ifmr} is equal to 0.13-0.14$M_{\odot}$; 
for $M_{\rm prog} $ around 2$M_{\odot}$ the range of WD masses is about 0.1$M_{\odot}$, whilst the differences vanish for initial masses around 1$M_{\odot}$.

In addition, we have also considered the solar-scaled theoretical IFMRs for [Fe/H]=0 (the typical metallicity of the semi-empirical IFMRs), $-$0.5 and $-$2.0, respectively, from the MIST evolutionary model calculations by \citet{bauermesa}. 
They are also displayed in Fig.~\ref{ifmr} (see also Table~\ref{tab}), which clearly shows the predicted effect of the initial progenitor metallicity on 
the IFMR according to \citet{bauermesa}. In general, at fixed $M_{\rm prog} $, the final 
$M_{\rm WD}$ tends to increase with decreasing progenitor metallicity. This trend is common to all theoretical IFMRs 
\citep[see, e.g.,][]{kl,wf,mmmb}, although, especially for $M_{\rm prog}$ below $\sim$2$M_{\odot}$, in some models the behaviour is non-monotonic \cite[see, e.g.][]{agbsynth}.
It is perhaps noteworthy that the solar-metallicity theoretical MIST IFMR (and also, e.g., the \citealt{kl, mmmb} theoretical IFMRs) is very close to the WEID IFMR, providing WD masses generally lower than the most recent semi-empirical IFMRs (apart from a narrow $M_{\rm prog}$ range). As described in the next section, we have used these theoretical IFMRs differentially to estimate the effect of using a metallicity-dependent IFMR for metal-poor globular clusters, instead of a roughly solar metallicity one, as is the case with the semi-empirical IFMRs.

Before closing this section, we want to mention an important and subtle issue. To calculate the initial CO stratification of our adopted WD models, the AGB evolution of a progenitor with a given initial total mass 
was stopped when the mass of its He-exhausted core --taken as the final WD mass-- had reached the appropriate value according to the 
CUMM IFMR (the same IFMR at all metallicities).
Changing the IFMR, in principle, changes the CO stratification of a given WD mass because of different He-core masses during the progenitor core He-burning phase, and different numbers of thermal pulses experienced during the AGB phase. This would require recalculating the initial CO profiles for all WD models whenever the IFMR is changed. 
In Appendix~B, we present numerical experiments which show how neglecting this recalculation should not affect the results of our analysis appreciably.

\section{Analysis}\label{S3}

As mentioned in Sect.~\ref{Section1}, we have explored three distinct age regimes, which we discuss separately. For each regime, we have considered the approximate age of a representative cluster among those we have 
studied in the past, and calculated the corresponding WD 
isochrones in the appropriate 
photometric filters (those used in our study of the representative cluster), one isochrone for each chosen IFMR.
To assess the differential effect of varying the IFMR under realistic observational conditions, the isochrones have been shifted in colours and magnitudes to account for the cluster extinction and distance modulus; in this way, we could account for realistic photometric errors (taken from 
photometric studies of the reference cluster) when computing 
the corresponding differential LFs using a Monte Carlo technique 
described e.g. in \citet{47Tucwdjwst}. 

The LFs have been calculated with the same bin size as the empirical LF of the reference cluster by including 50000 objects --to minimise statistical number fluctuations in the less populated bins-- then rescaled to have star counts 
roughly comparable to the observed LF of the reference 
cluster, and finally normalised to the same number of objects across a bright magnitude 
interval far from the age-sensitive portion of the LFs.

\subsection{Old ages ($\approx$10~Gyr)}
\label{old}

We have considered as representative of the old age regime, the 
case of the $\sim$12~Gyr old Galactic globular cluster 47~Tuc, which we have recently studied 
with both optical (HST) and infrared (JWST) data  \citep[see][]{47Tucwdjwst}. 

\begin{figure}
 \includegraphics[width=\columnwidth]{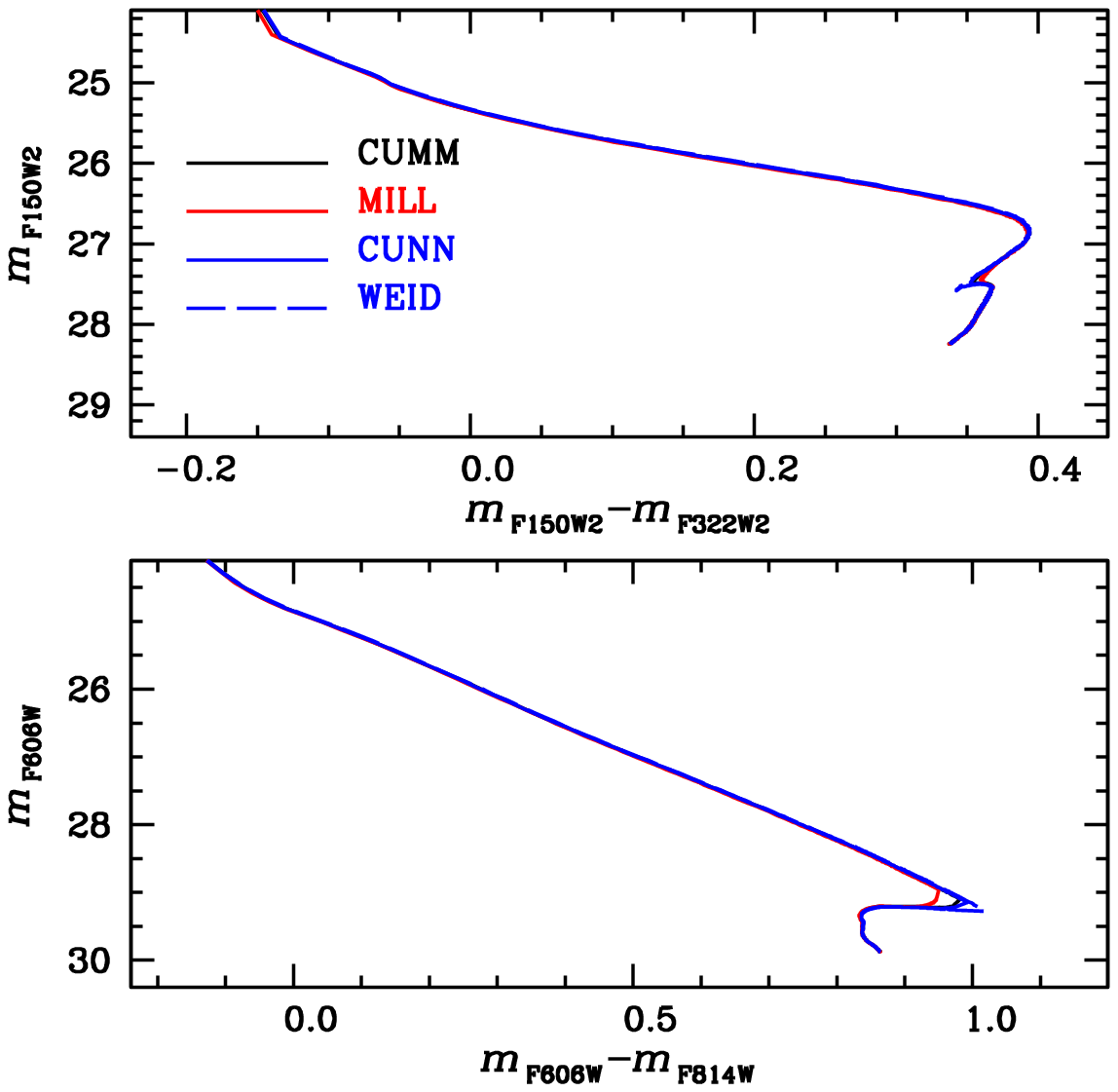}
 \caption{Four 12~Gyr WD isochrones for the chemical composition of 47~Tuc, displayed in the JWST/NIRCam (upper panel) and HST/ACS (lower panel) CMDs. The isochrones, displayed with the same colour codes of the upper panel of Fig.~\ref{ifmr}, have been calculated using the CUMM (solid black), CUNN (solid blue), MILL (solid red) and WEID (dashed blue) IFMRs, respectively (see text for details). 
 }
\label{isot12} 
\end{figure} 

\begin{figure}
\includegraphics[width=\columnwidth]{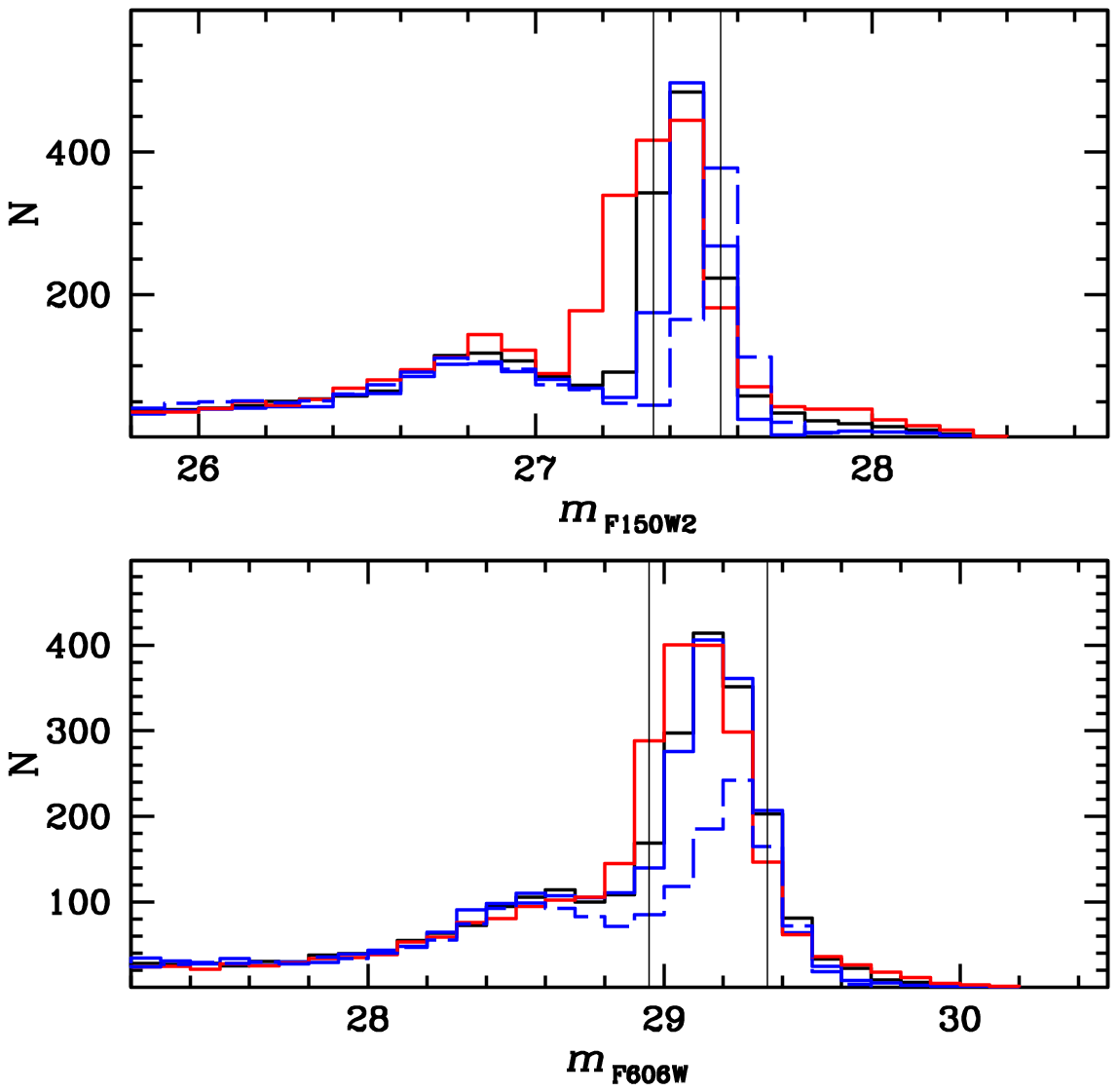}
 \caption{The LFs (0.1~mag bin size) calculated with the four isochrones of Fig.~\ref{isot12} (see text for details). The line styles and colour code are as in Fig.~\ref{isot12}.
 The thin vertical black lines mark the approximate magnitude range of the peak of the CUMM LFs, as a reference. The optical LFs are normalised to have the same 
 total number of objects at magnitudes $m_{\rm F606W}<$28.2, while the infrared LFs are normalised to the same total number of objects 
 at magnitudes $m_{\rm F150W2}<$26.4.
}
\label{isot12lf} 
\end{figure}

Figure~\ref{isot12} displays 12~Gyr WD isochrones \citep[calculated for the cluster metallicity, see][for details]{47Tucwdjwst} in the JWST/NIRCam and HST/ACS colour-magnitude diagrams (CMDs),    
computed with the CUMM, CUNN, MILL and WEID semi-empirical IFMRs. These isochrones are almost identical in the CMDs, except for very small, barely visible differences at the reddest, faintest end in the optical filters. 
The corresponding LFs for the bin size and photometric errors of the reference cluster data \citep[from][study]{47Tucwdjwst} are displayed in Fig.~\ref{isot12lf}.

\begin{figure}
\includegraphics[width=\columnwidth]{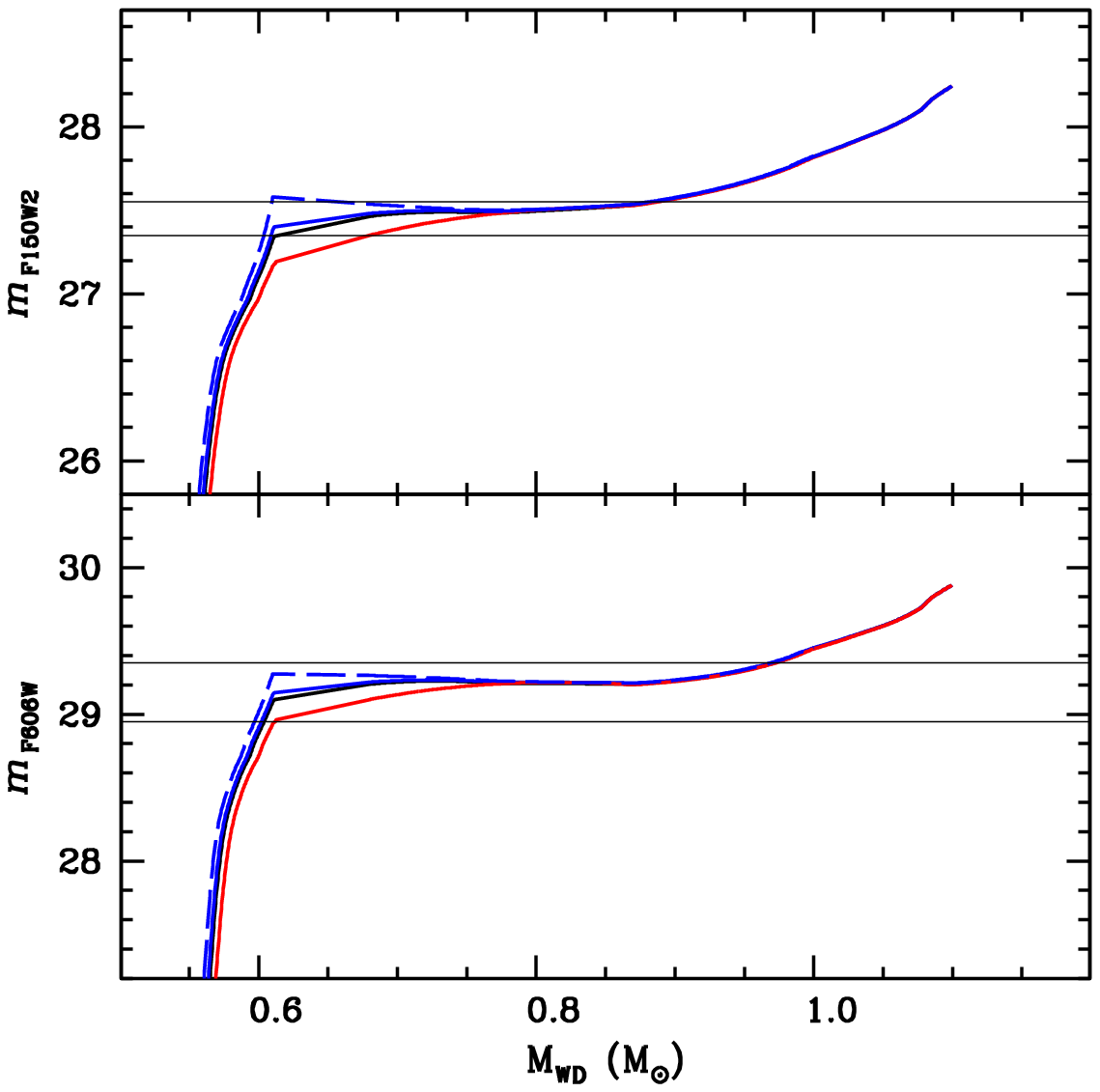}
 \caption{WD mass distribution as a function of the apparent F150W2 (upper panel) and F606W (lower panel) magnitudes for the isochrones of Fig.~\ref{isot12} (see text for details). The line styles are as in Fig.~\ref{isot12}.
 The thin horizontal black lines mark the approximate magnitude range of the peak of the LFs displayed in Fig~\ref{isot12lf}.
}
\label{isot12m} 
\end{figure}

In both HST and JWST filters, the four LFs show qualitatively the same shape (consistent with what is observed in old clusters), with a clear peak 
at their faint end, caused by the \lq{pile-up}\rq of increasingly massive WDs.
If we choose as a reference the calculations with the 
CUMM IFMR, the rising side of the peak of the MILL LF is brighter than the CUMM LF, whilst the opposite is true for the WEID LF, in both optical and infrared filters. The decreasing side  
of the peak and the termination of the LF agree amongst these three 
sets.  The CUNN LF is basically identical to the reference one.
This behaviour is explained by Fig.~\ref{isot12m}, which shows the WD mass distribution as a function of the apparent magnitude. 

Objects of near-constant mass populate the bright part of the isochrone 
and LF; they are the recently formed WDs, produced by the lower mass and longer-lived progenitors (of about 1$M_{\odot}$). This is easy to explain by remembering that along an isochrone, the sum of the WD cooling age ($t_{\rm cool}$) plus the lifetime of the progenitor ($t_{\rm prog}$) must be equal to the isochrone age ($t_{\rm iso}$). 
Given that bright WDs cool fast, $t_{\rm cool} \ll $ $t_{\rm prog}$, and $t_{\rm iso}\cong t_{\rm prog}$. This means 
that the progenitor, and as a consequence the WD mass, are roughly constant.
At fainter magnitudes, the WD cooling eventually slows, $t_{\rm cool}$ steadily increases and is no longer negligible compared to $t_{\rm prog}$; therefore, increasingly massive WDs (originating from more massive, shorter-lived progenitors) populate the isochrones.

Figure~\ref{isot12m} shows that in the HST optical filters, 
WDs with mass between $\sim$0.6 and $\sim$0.95$M_{\odot}$ 
populate a narrow magnitude range near the faint end of the sequence, producing the peak observed in the LF\footnote{See also the discussion in \citet{47Tucwdjwst} and \citet{hansen6397}.}. 
In the infrared, due to the behaviour of the bolometric corrections with respect to 
$T_{\mathrm eff}$ and surface gravity, the isochrones have a different shape compared to the optical, but still masses between $\sim$0.6 and $\sim$0.90$M_{\odot}$ are distributed across a narrow magnitude range and produce the observed peak of the LF. About 0.5-1.0~mag brighter than this faint peak, the LFs in Fig.~\ref{isot12lf} display either a less conspicuous local maximum, or a sharper increase followed by a plateau, depending on the photometric filter employed. This is caused by a slowing of the cooling of the $\sim$0.56$M_{\odot}$ models that populate the isochrone down to these luminosities. The slower cooling speed causes a first fairly sharp change of the rate of increase of the WD mass as a function of magnitude along the isochrone (see Fig.~\ref{isot12m}), which produces the mentioned feature in the LF \citep[this feature is confirmed empirically, see e.g.,][]{47Tucwdjwst}.

Increasing or decreasing the age of the isochrone decreases or increases the brightness of the peak in the LF (because of the increased/decreased cooling times, see Fig.~\ref{isot12app} in  Appendix~C), which is used as an age indicator 
\citep[see, e.g.][for some examples]{M4wd,N2158,M4wdjwst,47Tucwdjwst}. 
For this reason, in the following, we are most interested in studying how the brightness of the peak of the LF is affected by the choice of the IFMR. The actual individual values of the star counts are less relevant to our analysis because they depend on the exact number distribution of the WD masses along the isochrones, which in turn is affected by the choice of the progenitors' mass function and the cluster's dynamical evolution. 
Our LF calculations are with a fixed progenitor mass function, and different IFMRs alter the number distribution of WD masses along the isochrones indirectly, because they change the relationship between a given 
value of $M_{\rm prog}$ and the final mass $M_{\rm WD}$ (we will return to this point in the final section of the paper.

\begin{figure}
\includegraphics[width=\columnwidth]{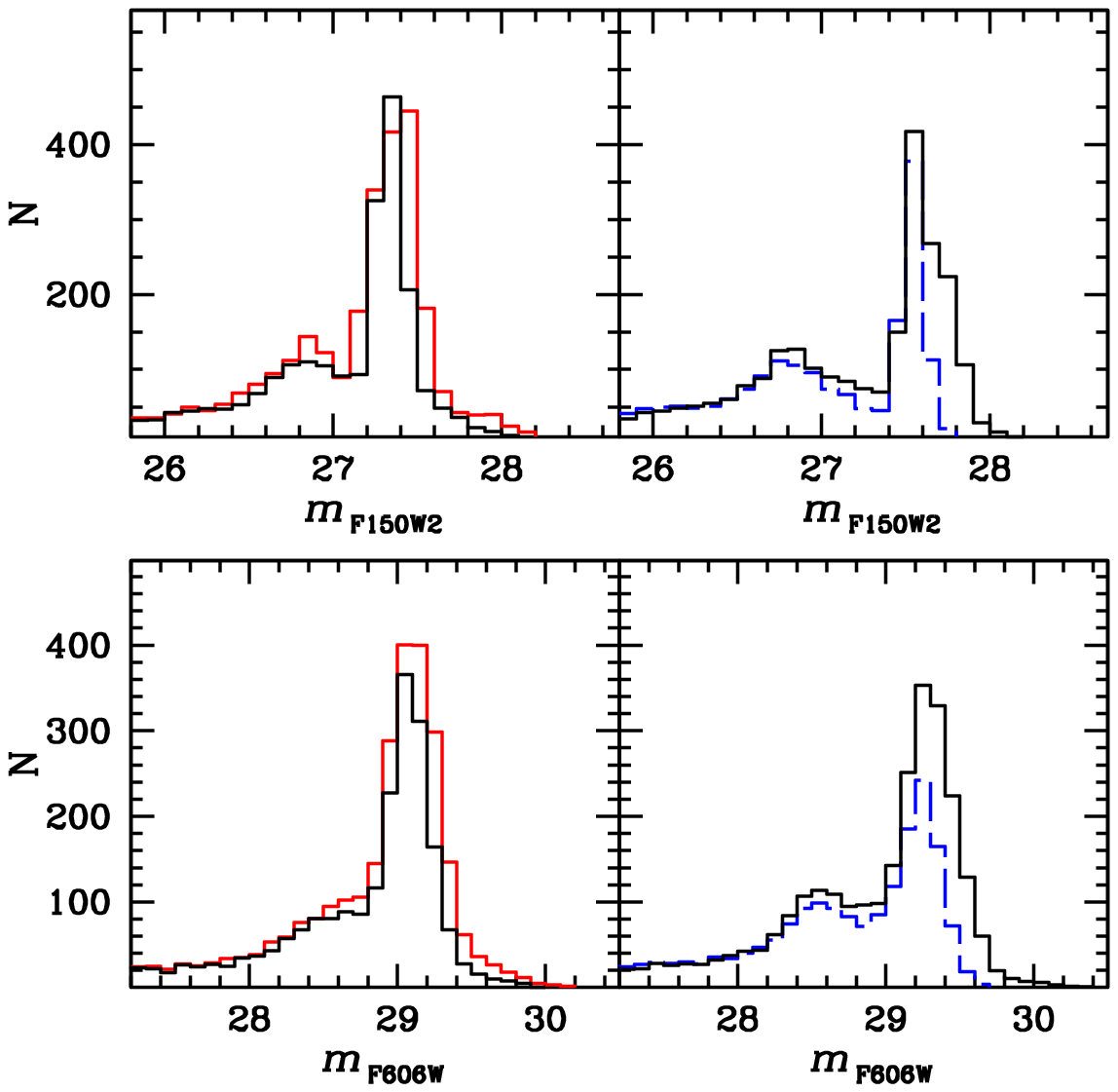}
 \caption{The left panels display a 11.4~Gyr CUMM LF and a 12~Gyr 
 MILL LF in the JWST (top) and HST (bottom) filters. The right panels 
 display a 12.4~Gyr CUMM LF and a 12~Gyr 
 WEID LF in the JWST (top) and HST (bottom) filters.
 Line styles and colour code are as in Fig.~\ref{isot12}.
 The LFs are normalised as in Fig.~\ref{isot12lf}.
}
\label{isot12multi} 
\end{figure} 

We have already observed that the brightness of the rising side of the LF peak is affected by the choice of the IFMR; this can be explained as follows.
At a fixed $M_{\rm WD}$=0.6$M_{\odot}$ and above, the progenitors for the MILL IFMR are less massive than the CUMM counterpart, whilst the opposite is true for the WEID IFMR, as shown by Fig.~\ref{ifmr}.
This means that, for a fixed isochrone age, WDs in this mass range have cooled down less in the case of the MILL IFMR, hence they are brighter compared to the CUMM isochrone. The opposite is true in the case of the WEID IFMR.
Differences are more substantial for $M_{\rm WD}$ between 0.6 and 0.7-0.75~$M_{\odot}$ stemming from lower mass progenitors. In this mass range, the variation of the progenitor mass due to the different choices of the IFMR causes a variation of $t_{\rm prog}$ on the order of a few 100~Myr, hence WD brightness variations on the order of 0.1-0.2~mag. At larger $M_{\rm WD}$ (above $\sim$ 0.7-0.75~$M_{\odot}$), the $t_{\rm prog}$ differences are up to one order of magnitude smaller, and the WD brightness variations become inappreciable, of just $\sim$0.01 magnitudes.
The effect of varying the IFMR is quantitatively only slightly larger 
in the infrared filters, in the mass range between $\sim$0.6 and $\sim$0.75$M{\odot}$.

This explains why the rising side of the LF peaks differ among the CUMM, WEID, and MILL isochrones. On the other hand, the CUNN LF is basically identical to the CUMM one because in the relevant 
$M_{\rm WD}$ range between 0.6 and 0.7-0.75$M_{\odot}$ the two IFMRs display minimal differences (see Fig.~\ref{ifmr}). 
Another consequence is that the LF calculated with the theoretical MIST IFMR for a solar progenitor metallicity is essentially identical to the WEID LF, because the two IFMRs overlap for $M_{\rm WD}$ values up to $\sim$0.8$M_{\odot}$ (Fig.~\ref{ifmr}).

Figure~\ref{isot12multi} shows that to match the magnitude of the 
rising side of the 12 Gyr MILL LF peak, the age of the CUMM LF  
needs to be 
decreased by 0.6~Gyr, while in the case 
of the 12 Gyr WEID LF, the age must be increased by 0.4~Gyr\footnote{The same is true for the CUNN LF.}. Variations of $\pm$0.2~Gyr around these values produce the same match, due to the width of the LF bins and the photometric errors (this 
also implies that the 12~Gyr CUNN and CUMM LFs agree within $\pm$0.2~Gyr).
Once the appropriate age shifts are applied, the entire magnitude range of the peaks of the MILL LFs is matched, not just the rising side. The same is not true for the WEID LF, due to the varying sensitivity of the brightness of WD models to different age variations. 

Finally, we have used the theoretical MIST IFMRs in Fig~\ref{ifmr} to address another important issue. When calculating WD LFs for globular clusters, the use of a semi-empirical IFMR based on systems at about solar metallicity implicitly assumes that the IFMR is metallicity-independent.
This is not necessarily true, and indeed theoretical stellar evolution models --keeping in mind all the associated uncertainties in the treatment of the AGB phase-- do predict a metallicity dependence. \footnote{We remark that the dependence of the IFMR on metallicity is an issue mainly for the old ages typical of globular clusters. The regimes of intermediate and young ages discussed in the next sections are typical for open clusters, which generally have metallicities around solar.}

We have employed the MIST IFMRs of Fig.~\ref{ifmr} differentially to have 
an estimate of the effect of metallicity on the WD LF, according to this particular set of stellar evolution models.
The IFMR for a scaled-solar [Fe/H]=$-$0.5 can be used in the case of a globular cluster like 47~Tuc ([Fe/H]$\sim -$0.7, [$\alpha$/Fe]$\sim$0.4), which has approximately the same total metallicity; the resulting 12~Gyr LF is practically identical to the 12~Gyr LF calculated with the [Fe/H]=0 IFMR, because the differences of the two IFMRs are negligible for WD masses below $\sim$0.7$M_{\odot}$.
Therefore, according to these MIST calculations, a solar IFMR is still appropriate for metal-rich globular clusters.

Let's consider as a reference a metal-poor globular cluster instead, like the $\sim$13~Gyr old cluster NGC~6397, which we investigated in \citet{n6397jwst} using JWST data. We have calculated WD isochrones as those employed in \citet{n6397jwst}, considering progenitors' lifetimes from $\alpha$-enhanced models with [Fe/H]=$-$1.9 appropriate for the cluster \citep[see][]{n6397jwst}, but using two different IFMRs: The MIST IFMR 
for solar composition (a proxy for the semi-empirical IFMRs), and the solar scaled [Fe/H]=$-$2 MIST IFMR. 
As mentioned, the cluster has [Fe/H]$\sim -$1.9 and an $\alpha$-enhanced metal distribution; therefore, its total metallicity is higher than the solar-scaled [Fe/H]=$-$2 composition of the IFMR, but it is still a reasonable approximation for our purposes. 

\begin{figure}
\includegraphics[width=\columnwidth]{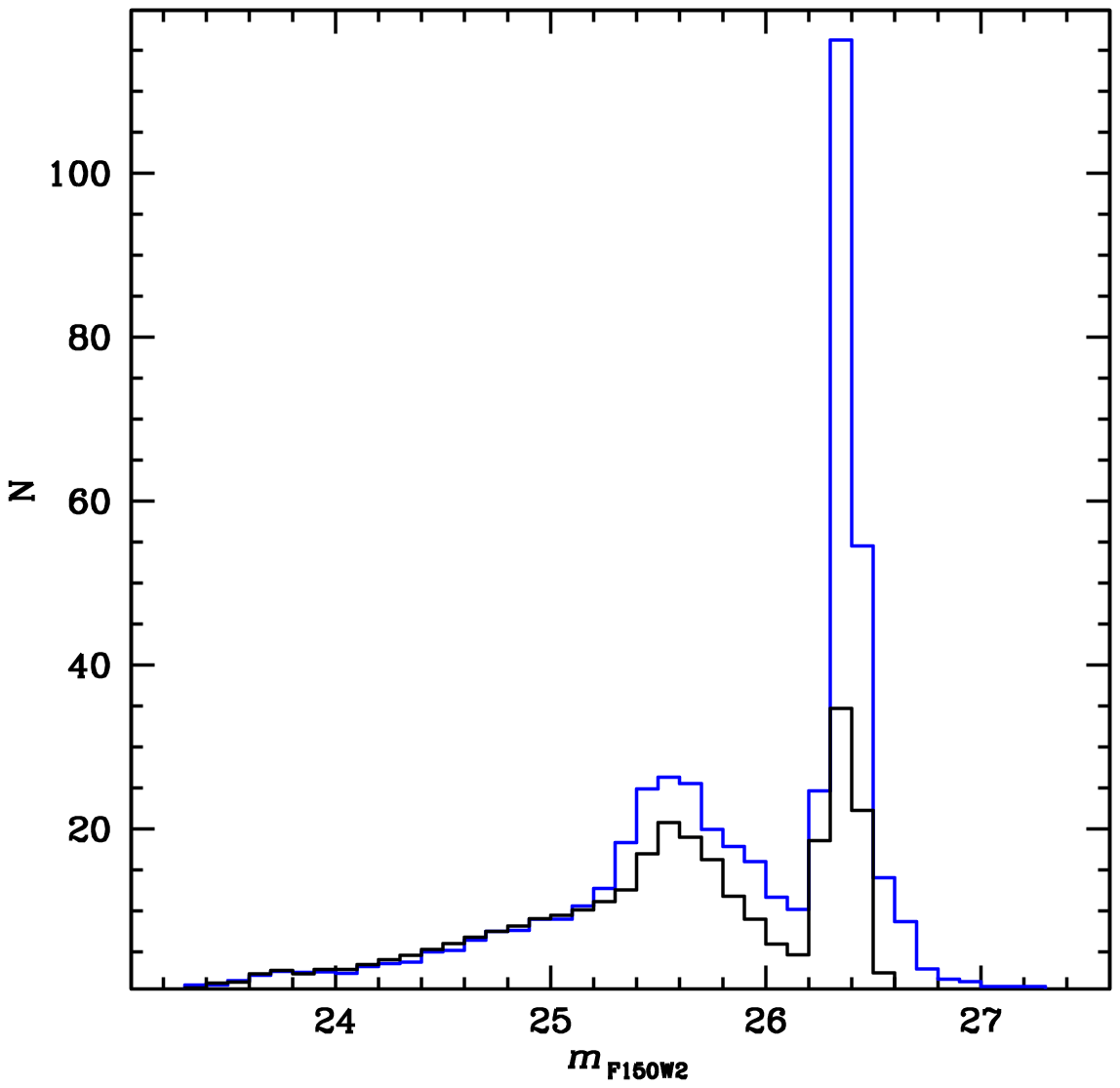}
 \caption{A 13~Gyr WD LF calculated for NGC~6397 metallicity and 
 the MIST theoretical solar scaled IFMR for [Fe/H]=$-$2 (blue line), together with a 12.2 Gyr WD LF calculated using the solar metallicity MIST IFMR (black line). The two LFs are normalised to the same total number of objects for $m_{\rm F150W} <$25.
}
\label{mpoorifmr} 
\end{figure}

Figure~\ref{mpoorifmr} shows that to match the magnitude range of the peak 
of the LF, the calculation with the solar metallicity IFMR has to be younger by 0.8~Gyr (with an uncertainty of $\pm$0.2~Gyr, again due to the bin size 
and photometric error). This means that the cluster age we obtain with a solar metallicity IFMR has to be increased  
by 0.8$\pm$0.2~Gyr to account for the effect of 
metallicity on the IFMR predicted by the MIST models.
We are probably slightly overestimating this difference, given that, as mentioned before, the cluster metallicity is not as low as that of the metal-poor IFMR employed. 

\subsection{Intermediate ages ($\approx$1~Gyr)}
\label{intermediate}

As representative of the intermediate-age regime, we have considered the $\sim$1.8~Gyr old open cluster NGC~2158, which we studied in \citet{N2158} using HST/ACS data. 

\begin{figure}
 \includegraphics[width=\columnwidth]{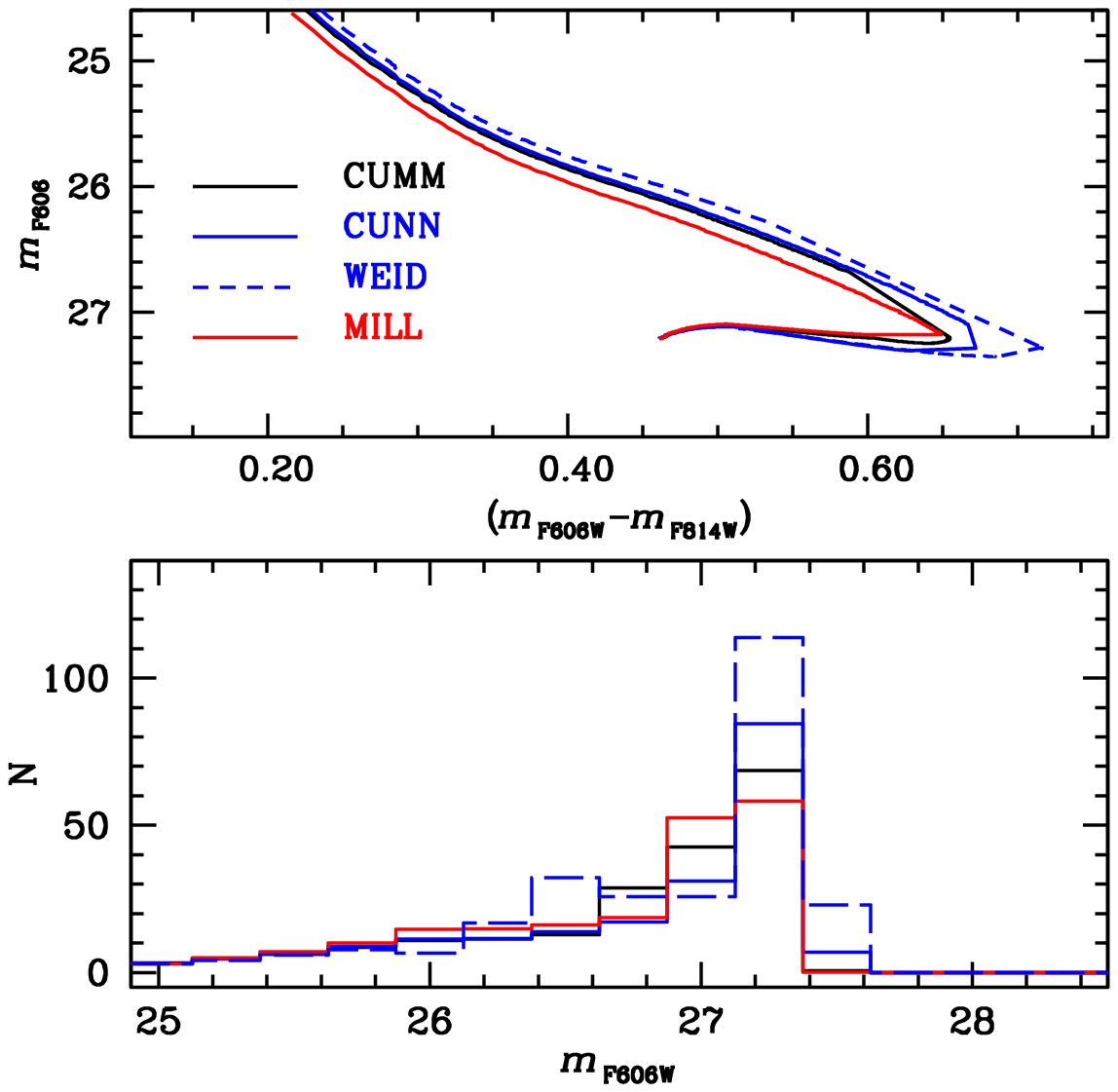}
 \caption{{\sl Upper panel:} Four 1.8~Gyr WD isochrones for the chemical composition of NGC~2158, displayed in the HST/ACS CMD. 
 The isochrones have been calculated using the CUMM (solid black), CUNN (solid blue), MILL (solid red), and WEID (dashed blue) IFMRs, respectively.
 {\sl Lower panel}: The LFs (0.25~mag bin size) calculated with the isochrones in the upper panel. The line styles are as in the upper panel.
The LFs are normalised to the same total number of objects 
 at magnitudes $m_{\rm F606W}<$26.4.
}
\label{isot18} 
\end{figure} 

Figure~\ref{isot18} shows 1.8~Gyr WD isochrones for roughly solar metal abundance \citep[calculated as in][]{N2158} in the HST/ACS colour-magnitude diagrams (CMDs), again computed employing the CUMM, CUNN, MILL, and WEID IFMRs, respectively, and the corresponding LFs. These have a 0.25~mag bin size and include the photometric errors from \citet{N2158} analysis; the bin size \citep[as in][]{N2158} is larger than in the previous globular clusters' case, to account for the less populated cooling sequences in open clusters. 

\begin{figure}
 \includegraphics[width=\columnwidth]{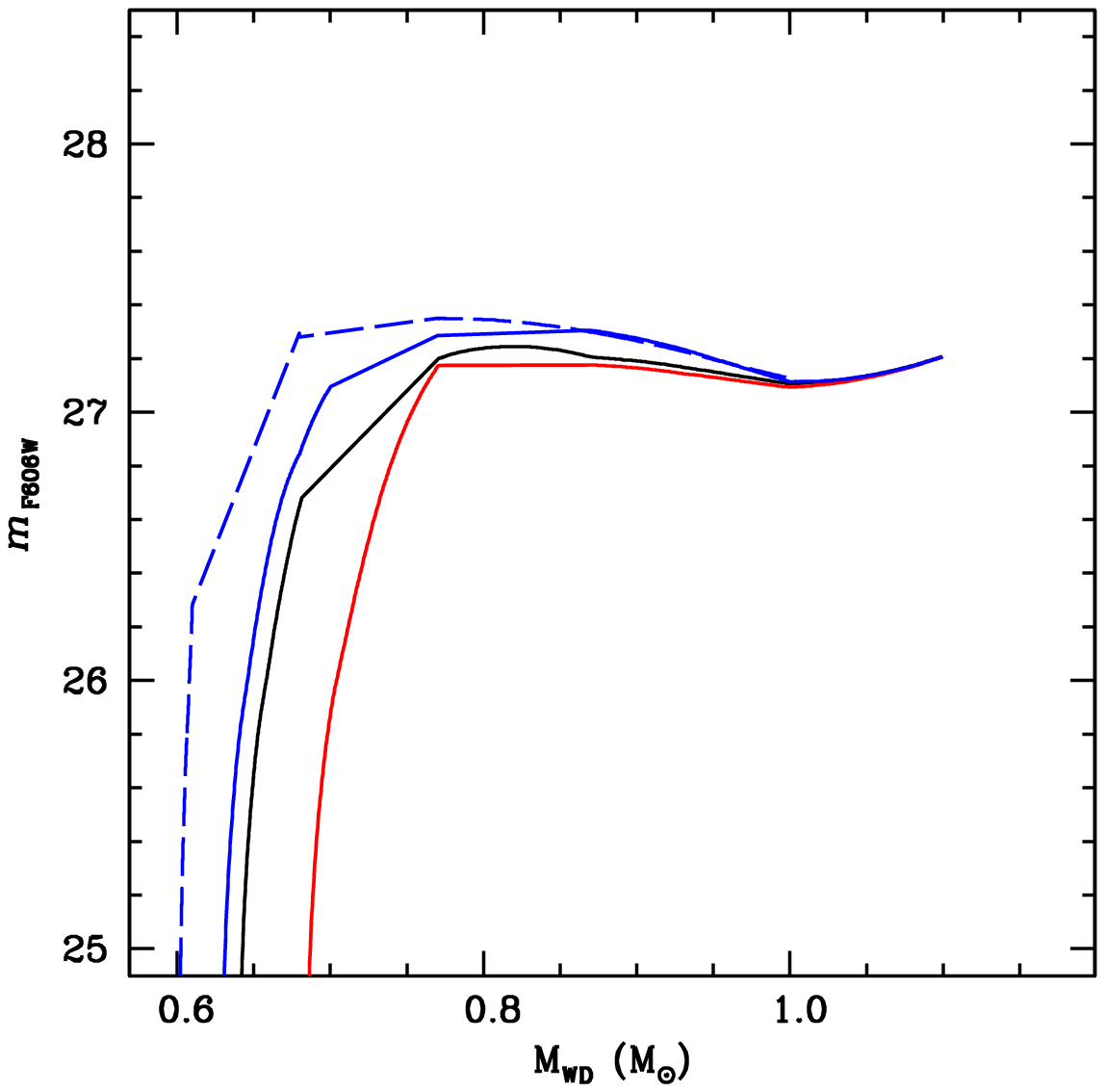}
 \caption{WD mass distribution as a function of the apparent F606W magnitudes for the isochrones of Fig.~\ref{isot18}. 
 The line styles are as in Fig.~\ref{isot18}.
}
\label{isot18mass} 
\end{figure}

The shape of the isochrones is qualitatively the same as for the old age range; they all start at bright magnitudes with an extended portion (in magnitude and colour) populated by objects with 
almost constant mass and decreasing brightness. 
The progenitor's mass for the brightest WDs at this age and chemical composition is $\sim$1.85$M_{\odot}$, and  
varying the IFMR causes the different 
colours at fixed magnitude (because of a different value of $M_{\rm WD}$ along the  
bright part of the isochrones). 
At their faint end, the isochrones turn towards progressively 
bluer colours due to the appearance of increasingly massive objects with decreasing radii, which are enclosed within a narrow magnitude range, as 
shown in Fig.~\ref{isot18mass}.
This 
produces a peak at the faint end of the corresponding LFs -- the age indicator, as in the old-age regime, see Fig.~\ref{isot18} and Fig.~\ref{isot18app} in Appendix~C-- as observed in NGC~2158 \citep[see][]{N2158}.

\begin{figure}
 \includegraphics[width=\columnwidth]{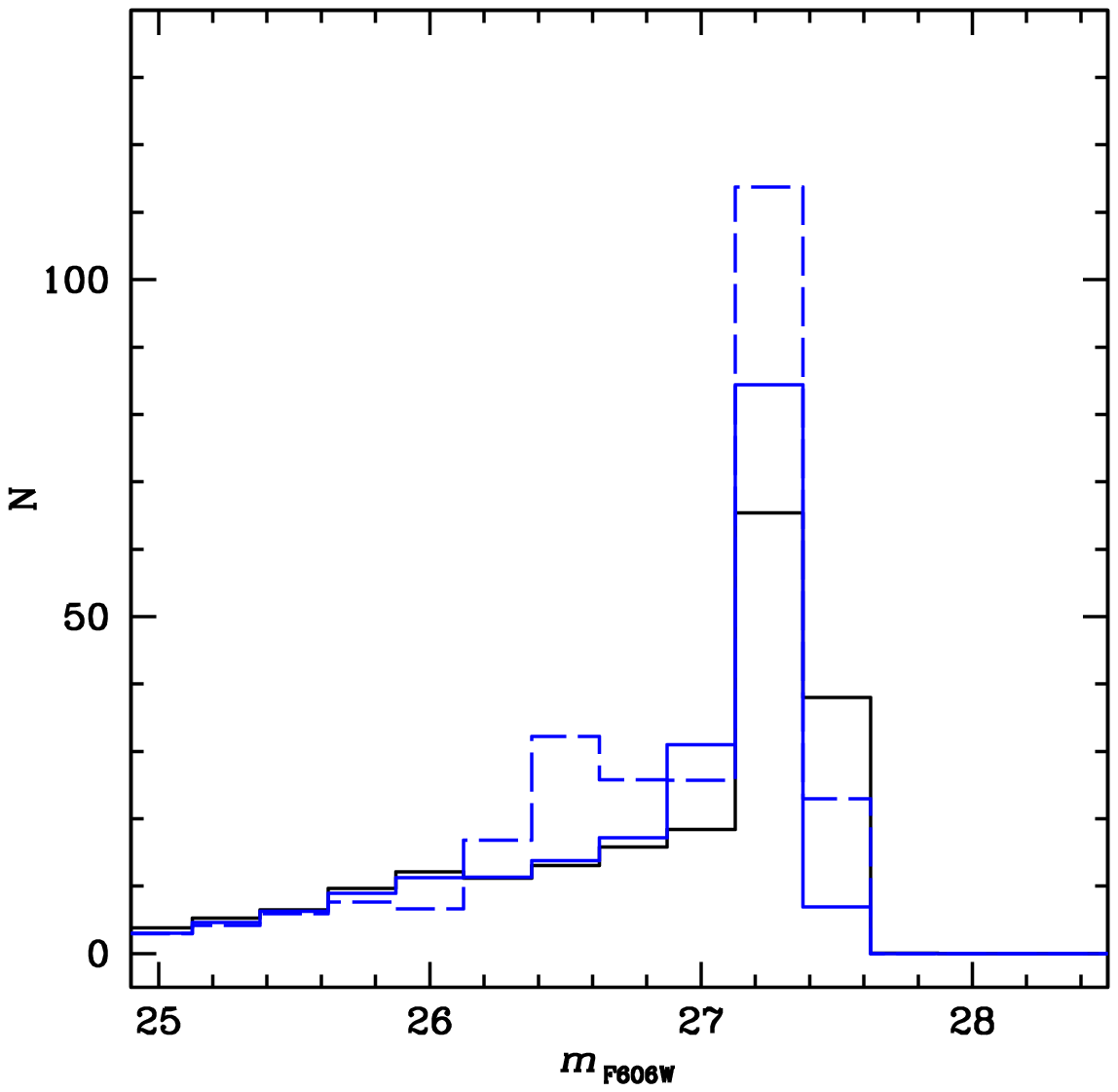}
 \caption{The 1.8~Gyr WEID (dashed blue) and CUNN (solid blue) LFs of Fig.~\ref{isot18},  together with a 2.0~Gyr CUMM LF (solid black)  -- see text for details).
 Line styles, colour code and normalisation are as in Fig.~\ref{isot18}.
}
\label{lft18test} 
\end{figure}

The brightness and width of the peak of the MILL and CUMM LFs are the same, whilst the CUNN and WEID LFs display a fainter tail, due to the fainter magnitude of the \lq{turn to the blue\rq} point of the corresponding isochrone (see Fig.~\ref{isot18}).
Figure~\ref{lft18test} 
shows that the age of the CUMM (and the MILL, not displayed for the sake of clarity) LF has to be increased by 200~Myr to match the magnitude of the faintest 
tail of the WEID and CUNN LFs (a variation of $\pm$100~Myr around this increase produces the same match).

\subsection{Young ages ($\approx$100~Myr)}
\label{young}

\begin{figure}
 \includegraphics[width=\columnwidth]{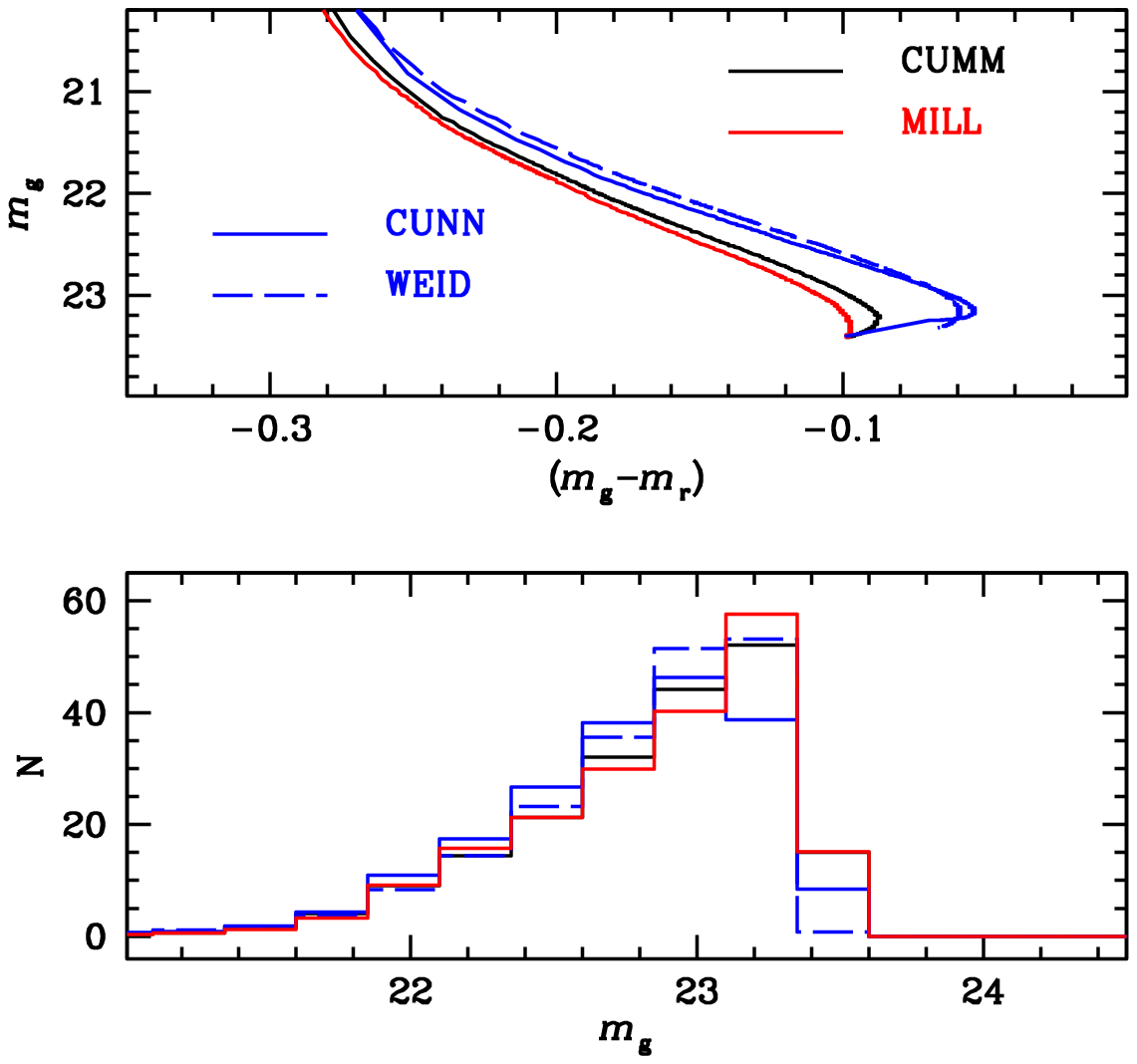}
 \caption{{\sl Upper panel:} Four 300~Myr WD isochrones for the chemical composition of M~37, displayed in the $Sloan$ CMD. 
 The isochrones have been calculated using the CUMM (solid black), CUNN (solid blue), MILL (solid red) and WEID (dashed blue) IFMRs, respectively.
 {\sl Lower panel}: The LFs (0.25~mag bin size) calculated with the isochrones in the upper panel. The line styles and colour code are also as in the upper panel.
The LFs are normalised to the same total number of objects 
 at magnitudes $m_{\rm g}<$22.
}
\label{isot300} 
\end{figure}

We have considered a 300~Myr age, 
approximately the same as the open cluster M~37, which 
we investigated in \citet{M37wdgaia} using $Sloan$ photometry. Figure~\ref{isot300}  shows the 300~Myr isochrones in $Sloan$ filters, calculated as in \citet{M37wdgaia} for the 
CUMM, WEID, CUNN and MILL IFMRs, respectively.
The corresponding LFs, including the photometric errors and bin size from our \citet{M37wdgaia} study, are also displayed. 
Their shape is different from that of older ages; after a flat part at the brightest magnitudes, star counts steadily increase, before a sharp drop at the faint end of the LF. 
The brightness of this drop is the age indicator, as shown in Fig.~\ref{isot300app} in Appendix~C.

\begin{figure}
 \includegraphics[width=\columnwidth]{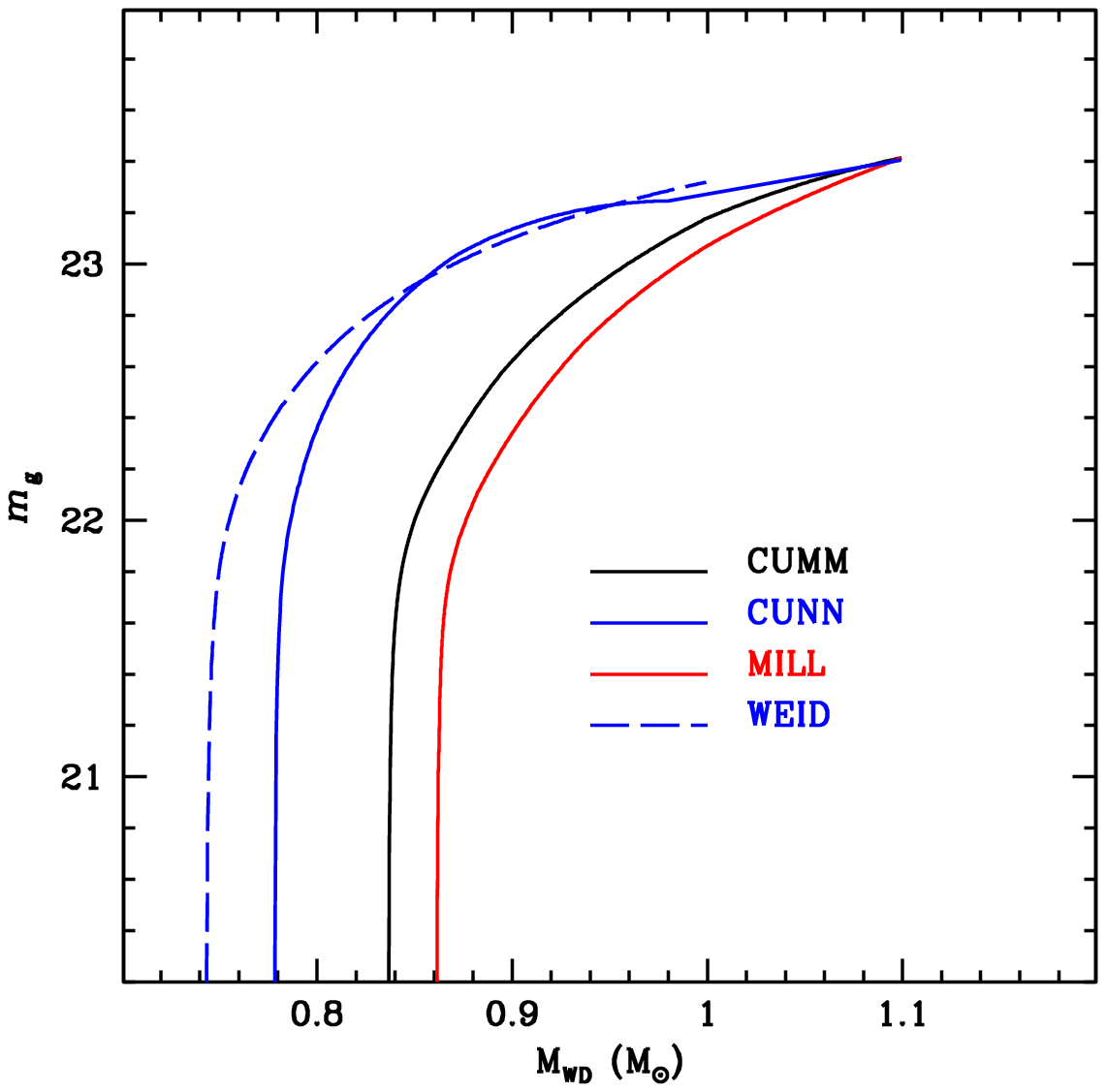}
 \caption{WD mass distribution as a function of the apparent $g$ magnitude for the isochrones of Fig.~\ref{isot300} (see text for details). The line styles are as in Fig.~\ref{isot300}.
}
\label{isot300m} 
\end{figure}

Figure~\ref{isot300m} displays the mass distribution as a function of the apparent $g$ magnitude for the isochrones of Fig.~\ref{isot300}. At this age, the value of $M_{\rm prog}$ for the newly formed WDs is $\sim$3.5~$M_{\odot}$,  and among the different isochrones the lowest mass WDs range between $\sim$0.74$M_{\odot}$ (for the WEID isochrone) and $\sim$0.86$M_{\odot}$ (for the MILL isochrone).
When the apparent $g$ magnitude increases above $\sim$22, 
$M_{\rm WD}$ begins to increase more steeply; there are no flat sections of the curves as for the older ages. Some flattening appears only for the CUNN isochrone due to the very different shape of the IFMR in the high $M_{\rm WD}$ regime -- 
see Fig.~\ref{ifmr}-- that reflects also on the shape of the isochrone (Fig.~\ref{isot300}).

The MILL, CUMM and CUNN LFs display a poorly populated tail at magnitudes fainter than the sharp drop, which is absent in the 
WEID LF. This is caused by the presence of objects with masses above 1.0$M_{\odot}$, the upper limit for the WEID IFMR. They are located across the faintest part of the isochrones and are absent in the WEID isochrone and LF. The age of the MILL, CUMM and CUNN LFs needs to be decreased by 50~Myr ($\pm$20~Myr) to deplete the faint tail of the LF populated by WDs more massive than 1.0$M_{\odot}$, and match the magnitude of the sharp termination of the WEID LF.

\section{Summary and discussion}\label{S4}

The previous section has shown how different IFMRs affect both the theoretical WD isochrones and LFs employed to determine the ages of star clusters. 

We found that the IFMR of WDs with mass between $\sim$0.6 and 0.7-0.75$M_{\odot}$ plays a major role at old (on the order of 10~Gyr) ages;  differences for higher WD masses do not affect the LF at these ages. 
When we consider the most up-to-date semi-empirical determinations of the IFMR derived from cluster and local field WDs --the CUMM, MILL and CUNN IFMRs-- and realistic observational conditions based on recent HST and JWST photometries, the age-sensitive feature of the WD LF (the magnitude of the peak at the faint end of the LF) is weakly affected by the choice of the IFMR. 
At a reference age of 12~Gyr, the LF computed with the latest IFMR based on local field WDs (CUNN) 
needs to be made younger by 0.6$\pm$0.2~Gyr to match the magnitude of the peak of a 12~Gyr MILL LF, which employs the most recent IFMR based on cluster WDs. The widely used CUMM IFMR based on cluster WDs gives the same results as the CUNN IFMR.
This systematic error source is quantitatively comparable to the current smallest formal errors on the WD ages of globular clusters, amounting to about $\pm$0.5~Gyr, like in the case of NGC~6752 \citep[][]{N6752hst}.

\begin{figure}
\includegraphics[width=\columnwidth]{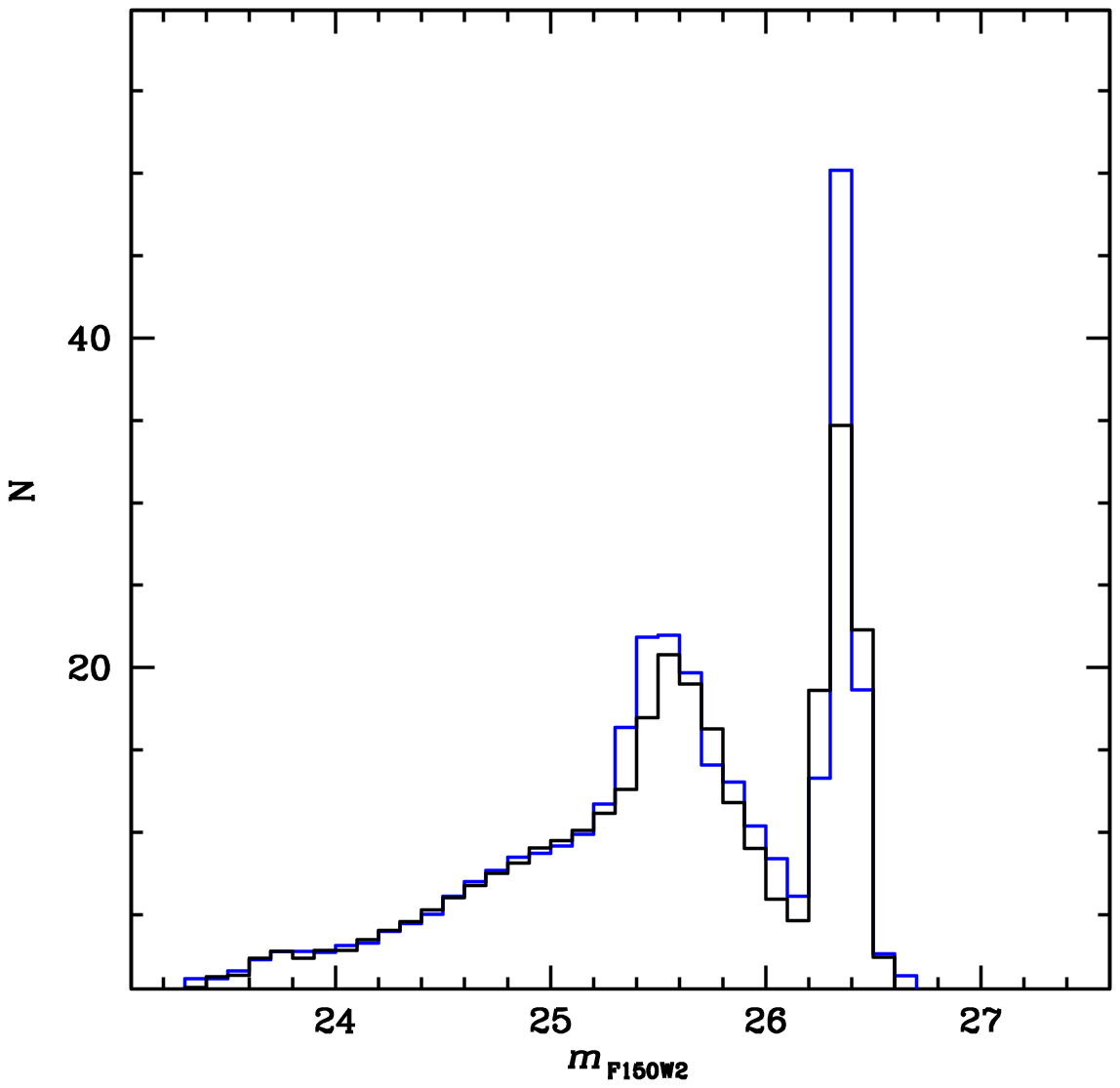}
 \caption{As Fig.~\ref{mpoorifmr}, but the 
 13~Gyr WD LF with the MIST IFMR for [Fe/H]=$-$2 (blue line)  calculated with a bottom-heavy mass distribution of the progenitors (black line -- see text for details).
}
\label{mpoorifmrimf} 
\end{figure} 

The effect of the progenitor metallicity on the IFMR is potentially slightly more important for metal-poor globular clusters. To date, we can only use theoretical models --with the associated uncertainties in the treatment of the AGB evolution-- to explore this issue comprehensively. According to the IFMRs from recent MIST stellar models, ages older by 
up to $\sim$0.8$\pm$0.2~Gyr would be obtained for metal-poor globular clusters when the IFMR for the appropriate progenitor metallicity is used instead of an IFMR for solar metallicity, the approximate metallicity of the populations on which the semi-empirical IFMRs are based-- see Fig.~\ref{mpoorifmr}. 
A high-precision, as homogeneous as possible determination 
and comparison of the MS turn-off and WD CS ages --same stellar models for the WD progenitors and for the isochrones used to determine the MS ages-- for the metal poor globular cluster NGC~6397\footnote{NGC~6397 is the most metal-poor globular cluster with photometry that reaches the faint end of its cooling sequence} and the metal rich one 47~Tuc will be necessary to search for any metallicity-dependent mismatch between the two determinations, beyond the associated errors. 
If the mismatch (if any) goes beyond the predicted effect of the uncertainty on the semi-empirical IFMRs discussed before, that would be a potential signature of composition systematics in the IFMR, which can then be compared with theoretical predictions.

Let's step back again to Fig.~\ref{mpoorifmr} to discuss briefly an issue previously mentioned in Sect.~\ref{S3}. 
The two LFs display the same magnitude of the peak at their faint end, the same general shape, but very different actual star counts, even after the normalisation. This could, in principle, allow us to discriminate between the IFMRs when fitting to real data, if the errors on the observed star counts are small enough.
In practice, this is not straightforward to achieve because, for a fixed IFMR, to predict the exact number distribution of $M_{\rm WD}$ in the observed field, it is necessary to have an accurate dynamical model of the target cluster, perhaps aided by observations of the WD and MS LFs at different locations over a large cluster area. The reason is that stellar dynamics may have modified over time the spatial distribution of the WDs in different ways, depending on their mass. A $M_{\rm WD}$ number distribution based on a fixed progenitor initial mass function may not 
be the appropriate one for the target cluster. The dynamical evolution can, in principle, enhance or minimise the differences seen in Fig.~\ref{mpoorifmr}.

As a simple numerical experiment to highlight this point, we have 
changed the progenitor initial mass distribution of one of the LFs in 
Fig.~\ref{mpoorifmr}, as a proxy for a change in the WD mass distribution due to the cluster dynamical evolution.
Figure~\ref{mpoorifmrimf} is as Fig.~\ref{mpoorifmr}, but this time the 13~Gyr LF for [Fe/H]=$-$2 IFMR has been calculated using a bottom-heavy number distribution of progenitor masses (exponent $k$=$-3.5$), giving more weight to lower-mass WDs along the isochrone. 
The two LFs are now also much more similar in star counts across the entire magnitude range, not just in the same general shape and magnitude of the peak at their faint ends. 

Moving to intermediate ages (on the order of 1~Gyr), the differences amongst the CUMM, CUNN and MILL IFMRs have an impact at the level of 
at most 0.2$\pm$0.1~Gyr for a reference age of 1.8~Gyr. 
Also in this case, the CUNN LF need to be made younger (by 0.2$\pm$0.1~Gyr) to match the magnitude of the age-sensitive peak of the MILL LF. In this age regime, the CUMM IFMR is in agreement with the MILL LF.
It is interesting here to refer to the work by \citet{barrientos},  
who studied a sample of wide double WD systems and, among others, they determined the effect of changing the adopted IFMR from CUMM to CUNN (keeping unchanged the CO profiles of the WD models, as in our analysis, see Sect.~\ref{S2} and Appendix~B) on the WDs total ages (cooling age plus progenitor lifetime). The total ages of their sample are typically of a few Gyr, and they found a variation of $\sim$0.36~Gyr, in broad agreement with our results for intermediate ages.

At young ages (on the order of 100~Myr), the shape of the WD LF changes. In the optical, there isn't a peak at the faint end of the LF, as for old and intermediate ages. The magnitude of the faintest bin of the LF is the age indicator, and the effect of changing the IFMR among the MILL, CUMM and CUNN results is negligible. At these ages, the maximum value of $M_{\rm WD}$ in the IFMR for CO WDs plays a role in determining the magnitude of the faintest bin (see Fig.~\ref{isot300m}), whilst it is not crucial at old and intermediate ages 
(the peak of the LF is not populated by the highest mass WDs, as shown e.g. in Figs.~\ref{isot12m}).

\begin{acknowledgements}
We thank our anonymous referee for a very constructive report, which has helped us improve the paper.
S.C, acknowledges financial support from INAF Theory Grant (PI: S. Cassisi) and INAF Science Network grant (PI: M. Marconi).
\end{acknowledgements}

\bibliographystyle{aa}
\bibliography{main.bib}

\begin{appendix}
\label{appa}
\section{The impact of ONe-core WDs}
As mentioned in Sect.~\ref{S2}, in our analysis, we did not include WD models with mass above 1.1$M_{\odot}$, which are predicted 
to have ONe cores and are not available in our adopted model database \citep{bastiwd}. We estimate here the impact of these more massive 
WDs on our results, considering the ONe-core models 
by \citet{camisassa} with hydrogen atmospheres, and masses between 1.1 and 1.29$M_\odot$.

In the old- and intermediate-age regimes, the ONe-core models reach luminosities that are much fainter (by one magnitude or more) than the faint end of the CO-core WD isochrones. Their neglect does not affect the results of our analysis based on the faint end of the CO-WDs LF, which is what is actually observed in star clusters of these ages. 

In the young age regimes, for ages up to $\sim$400-500~Myr, the ONe-core models are actually brighter than the faint end of the CO-core WD LF. This implies that our results based on the magnitude of the faint end of the CO-core WD LF should be unaffected.
In the narrow age range between $\sim$500 and $\sim$800-900~Myr the ONe-core WDs transition from being brighter to becoming fainter than the faintest CO-core WDs, and only in this regime, the inclusion of ONe-core WDs in the analysis would be necessary.

\section{Effect of changing the IFMR on the CO-stratification}
\label{appb}
In Sect.~\ref{S2} we mentioned that the chemical stratification of a WD with mass $M_{\rm WD}$ depends on the 
progenitor mass, hence on the adopted IFMR. In our analysis, we kept the set of WD models fixed when we changed the IFMR, and here we present numerical experiments to assess how this inconsistency may affect our results.

\begin{figure}
\includegraphics[width=\columnwidth]{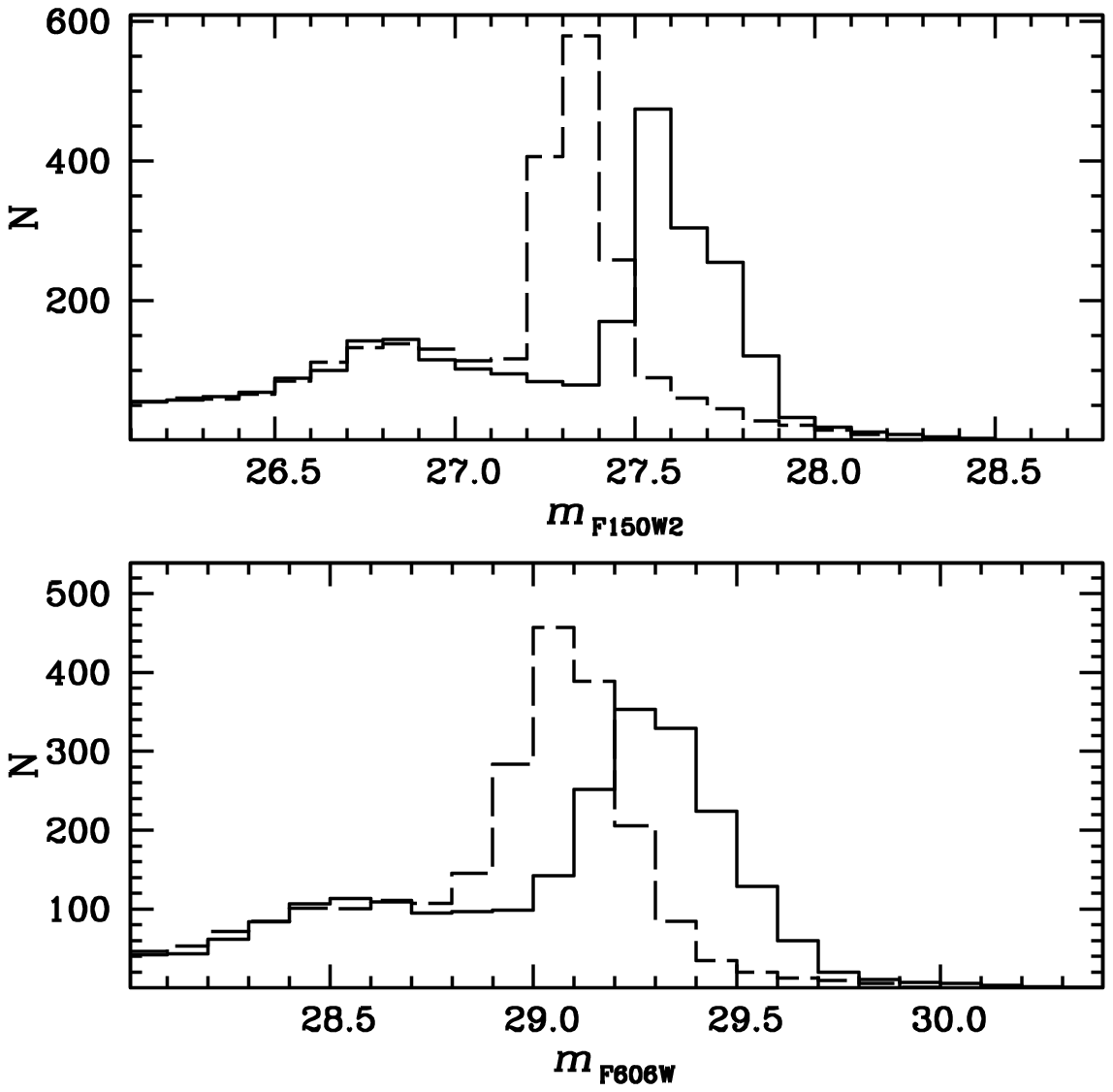}
 \caption{Two CUMM LFs for 11.4~Gyr (dashed line) and 
 12.4~Gyr (solid line), respectively (see text for details).
}
\label{isot12app} 
\end{figure}

For the old-age regime (see Sect.~\ref{old}), we considered the 0.61$M_{\rm WD}$ model as a proxy for all WD masses that lie in the magnitude range of the peak of the LF, the age indicator (see Figs.~\ref{isot12lf}, \ref{isot12m}, and \ref{isot12app}). 
For this value of $M_{\rm WD}$, the extremes of the variations of the progenitor mass with respect to the CUMM IFMR (the reference for our adopted WD models) are equal to $-0.2~M_{\odot}$ for the MILL IFMR, and $+$0.8~$M_{\odot}$
for the WEID IFMR. In the case of the MIST theoretical IFMRs, the variation of the progenitor mass 
between the solar and the [Fe/H]=$-$2 IFMRs is equal to $-$0.7$M_{\odot}$.
To maximise the effect on the CO profiles, we have therefore calculated CO profiles as in \citet{bastiwd} but varying the progenitor mass by $\pm$1.0$M_{\odot}$ around the reference value from the CUMM IFMR (see Sect.\ref{S2}); we found that for cooling ages on the order of 10~Gyr the 
0.61$M_\odot$ model cooling times at fixed magnitude vary at most 
by about only $\mp$140~Myr. 

\begin{figure}
\includegraphics[width=\columnwidth]{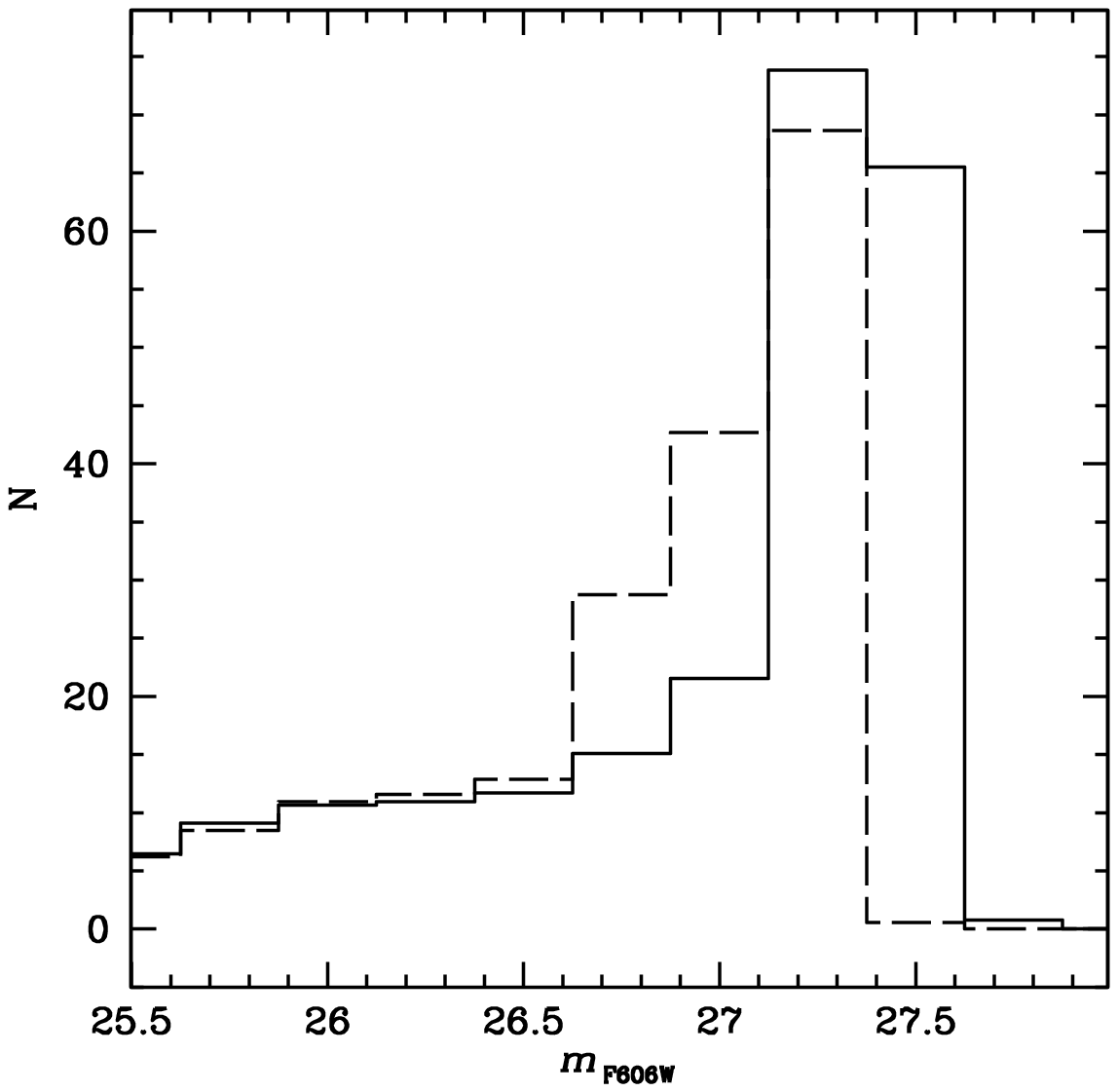}
 \caption{Two CUMM LFs for 1.8~Gyr (dashed line) and 
 2.1~Gyr (solid line), respectively (see text for details).
}
\label{isot18app} 
\end{figure}

\begin{figure}
\includegraphics[width=\columnwidth]{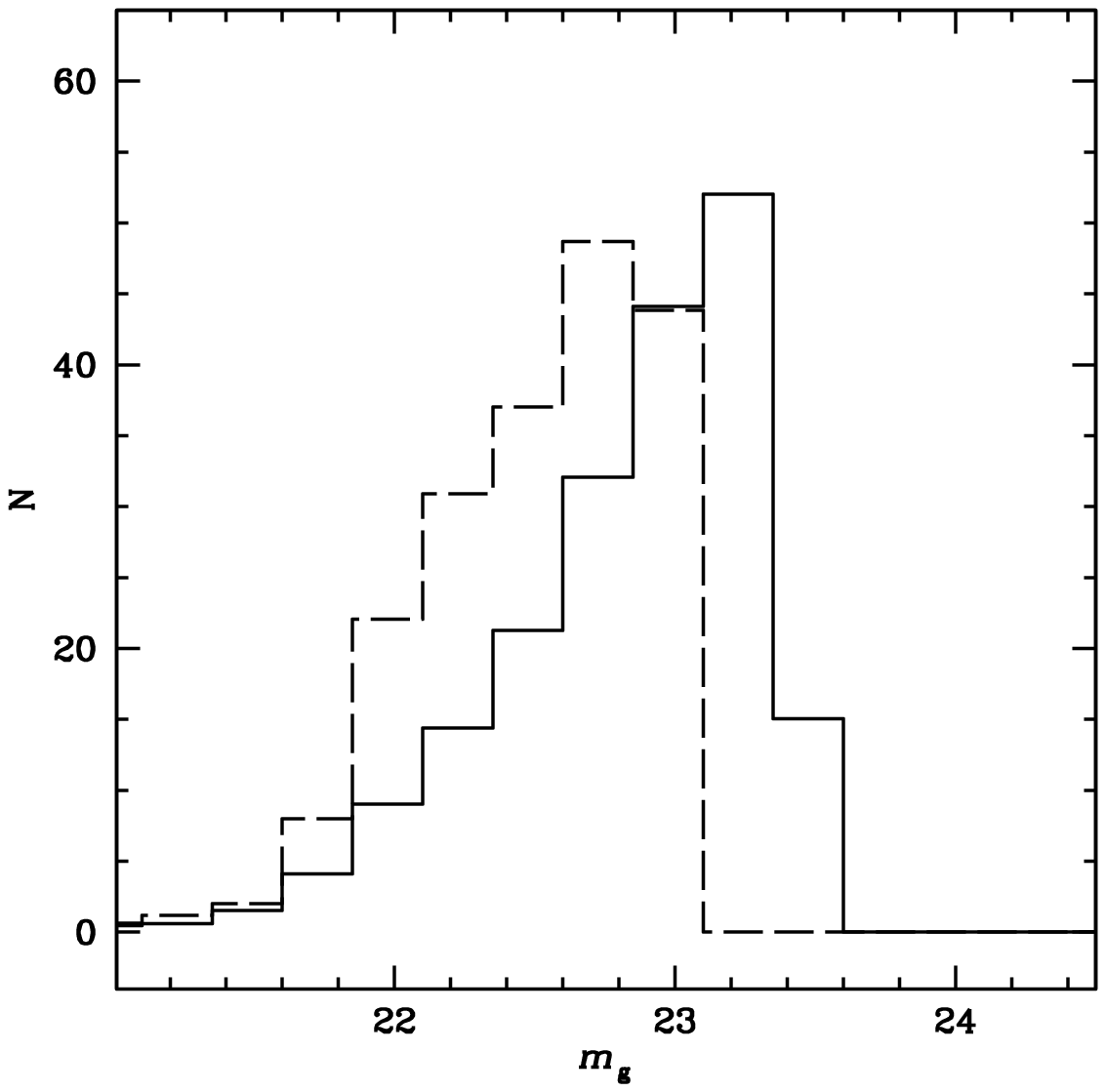}
 \caption{Two CUMM LFs for 200~Myr (dashed line) and 
 300~Myr (solid line), respectively (see text for details).
 }
\label{isot300app} 
\end{figure}

For the intermediate-age case, we considered the 1.0~$M_{\odot}$ model as a proxy for the WD masses in the magnitude range of the peak of the LF (see Figs.~\ref{isot18}, \ref{isot18mass}, and \ref{isot18app}). This model is also appropriate for studying the young-age regime because its mass is close to the LF faint cut-off magnitude (see Figs.~\ref{isot300}, ~\ref{isot300m}, and ~\ref{isot300app}).

 For $M_{\rm WD}$=1.0~$M_{\odot}$, the variations of the progenitor mass with respect to the CUMM IFMR are equal to $-0.6~M_{\odot}$ for the MILL IFMR, $+$0.5~$M_{\odot}$ for the CUNN IFMR, and $+$1.5 $M_{\odot}$ for the WEID IFMR, respectively. We have therefore 
 calculated CO profiles varying the progenitor mass by +$1.5~M_{\odot}$ and $-$0.5$M_{\odot}$ around the reference CUMM 
 value; we found that for cooling ages up to 8-9Gyr, the impact 
 of the different CO-profiles on the cooling times at fixed magnitude is negligible, 
 reaching at most $\sim$20~Myr for the older ages, when the progenitor mass is varied by $+$1.5 $M_{\odot}$.

\section{Additional figures}
\label{appc}
This appendix contains three additional figures discussed in Sect.~\ref{S3}, which show the age-sensitive features of WD LFs in different age regimes.

\end{appendix}
\end{document}